\RequirePackage{fix-cm}
\documentclass[twocolumn,epjc3]{svjour3}
\smartqed  % flush right qed marks, e.g. at end of proof
\usepackage{amsmath}
\usepackage{amssymb}
\usepackage{amssymb,bbold}
\usepackage{kantlipsum,widetext}
\usepackage{subcaption}
\usepackage{graphicx}
\RequirePackage{graphicx}
\usepackage{float}
\RequirePackage[colorlinks,citecolor=blue,urlcolor=blue,linkcolor=blue]{hyperref}
\RequirePackage[numbers,sort&compress]{natbib}

\journalname{Eur. Phys. J. C}
\begin{document}

\title{Charged scalar bosons in a Bonnor-Melvin-$\Lambda$ universe at conical approximation%\thanksref{t1}
}

\newcommand\correspondingauthor{\thanks{E-mail: \href{mailto:aobispo@utp.edu.pe}{aobispo@utp.edu.pe} (corresponding author)}}
\author{Luis B. Castro\thanksref{e1,addr1} \and Angel E. Obispo\thanksref{e2,addr2,addr3} \and Andr\'{e}s G. Jir\'{o}n\thanksref{e3,addr2} %etc.
}

\thankstext{e1}{E-mail: \href{mailto:luis.castro@ufma.br}{luis.castro@ufma.br}}
\thankstext{e2}{E-mail: \href{mailto:aobispo@utp.edu.pe}{aobispo@utp.edu.pe} (corresponding author)}
\thankstext{e3}{E-mail: \href{mailto:E20084@utp.edu.pe}{E20084@utp.edu.pe}}


\institute{Departamento de F\'{\i}sica, Universidade Federal do Maranh\~{a}o (UFMA), Campus Universit\'{a}rio do Bacanga, 65080-805, S\~{a}o Lu\'{\i}s, MA, Brazil \label{addr1} \and
Universidad Tecnol\'{o}gica del Per\'{u} (UTP), Los Olivos, Lima, Per\'{u} \label{addr2}\and
Universidad Privada del Norte (UPN), Los Olivos, Lima, Per\'{u} \label{addr3}
}

\date{Received: date / Accepted: date}
\maketitle
\begin{abstract}
The quantum dynamics of charged scalar bosons in a Bonnor-Melvin-$\Lambda$ universe is considered. In this study, the behavior of charged scalar bosons is explored within the framework of the Duffin-Kemmer-Petiau (DKP) formalism. Adopting a conical approximation ($\Lambda\ll 1$), we are considered two scenarios for the vector potential: a linear and quadratic vector potentials. In particular, the effects of this background in the equation of motion, phase shift, $S$-matrix, energy spectrum and DKP spinor are analyzed and discussed. 

\end{abstract}

\section{Introduction}
\label{intro}

The first-order Duffin-Kemmer-Petiau (DKP) formalism \cite{Petiau1936,Kemmer1938,PR54:1114:1938,Kemmer1939}
is employed to describe spin-zero and spin-one particles. It has been utilized to examine relativistic interactions between spin-zero and spin-one hadrons and nuclei, providing an alternative to their conventional second-order counterparts, the Klein-Gordon (KG) and Proca equations. While these formalisms are equivalent in the case free and involving minimally coupled vector interactions \cite{PLA244:329:1998,PLA268:165:2000,PRA90:022101:2014}, it is worth emphasizing that the DKP formalism offers a broader range of coupling possibilities that cannot be expressed within the confines of the Klein-Gordon (KG) and Proca theories \cite{PRD15:1518:1977,JPA12:665:1979}. 

Solutions of the DKP equation in curved space-time have been obtained for many systems and have been studied extensively in the literature in recent years \cite{EPJC44:287:2005,IJMPA25:2747:2010,AP343:40:2014,EPJP130:236:2015,EPJC75:287:2015,EPJC76:61:2016,EPJP132:541:2017,EPL118:10002:2017,
EPJC78:93:2018,GRG50:104:2018,IJMPA34:1950056:2019,IJMPA34:1950082:2019,GRG52:25:2020,IJMPA35:2050107:2020,MPLA36:2150059:2021,IJMPE30:2150050:2021,
CQG39:075007:2022,PS98:065224:2023,FBS64:13:2023}. In these references, the authors are considered various models of curved space-time and investigated the effects of such backgrounds on the energies of bound states of DKP particles. These investigations are significant as they explore the relationship between curved space-time and relativistic particle dynamics. They provide valuable insights into the fundamental nature of the universe by studying the behavior of particles in extreme gravitational environments such as black holes and cosmological structures. 

Magnetic fields are crucial in understanding various astrophysical phenomena across different distance scales, from stars to intergalactic regions. They shape celestial dynamics, influence accretion disks around black holes, and regulate processes in galactic nuclei. Even on large scales, magnetic fields impact cosmic matter distribution. Incorporating the magnetic field into the metric is not a trivial matter, but some solutions have been proposed. Notable examples include the Bonnor-Melvin metric \cite{PPSA67:225:1954,PL8:65:1964}, which describes a magnetic universe with a magnetic field aligned in the $z$-direction, and the Gutsunaev solution \cite{PLA123:215:1987,PLA132:85:1988}, which refers to a magnetic dipole. A theoretical description of fermions and bosons in a Bonnor-Melvin metric has been addressed in recent literature \cite{EPJC76:560:2016,ADHEP2018:1953586:2018}. In these work, fermions and scalar bosons are described by the Dirac and Klein-Gordon (KG) formalisms, respectively. Currently, this is an open problem that deserves further exploration. 

The Bonnor-Melvin universe, as presented in references \cite{PPSA67:225:1954,PL8:65:1964}, represents an exact solution to the Einstein-Maxwell equations. This solution characterizes a static, cylindrically symmetric (electro)magnetic field immersed within its own gravitational field. The magnetic field is aligned with the symmetry axis, but it is not homogeneous. Recently, as demonstrated in \cite{PRD99:044058:2019}, the author extended the Bonnor-Melvin solution to encompass a nonvanishing cosmological constant (Bonnor-Melvin-$\Lambda$ universe). In this setting, the space-time maintains its cylindrical symmetric and static nature, but, unlike the original solution, it genuinely represents a homogeneous magnetic field (intrinsic). A theoretical description of scalar bosons in a Bonnor-Melvin-$\Lambda$ universe has been addressed in \cite{barbosa2023scalar}. The authors are solved the KG equation and determined the Landau levels for the case free and for a Coulomb-like scalar potential. However, the investigation of scalar bosons in a Bonnor-Melvin-$\Lambda$ universe using the DKP formalism remains uncharted territory. Hence, we believe that this unexplored problem holds significant potential for further study.

Inspired by these studies, we explore the dynamics of a charged scalar boson using the DKP formalism immersed in a Bonnor-Melvin-$\Lambda$ universe. Our analysis examines the effects of both the intrinsic magnetic field and an additional external magnetic field associated with a four-vector potential $A_{\mu}$, which is minimally coupled. We aim to examine the quantum effects associated with gravitation from the perspective of relativistic quantum mechanics in curved spacetime. To achieve this, we analyze the scalar sector of the DKP equation in this context and derive the equation of motion for an arbitrary form of the vector potential. Given the complexity of the equation of motion, we employed the approximation $\sqrt{2\Lambda}r \ll 1$, or equivalently, $\Lambda \ll 1$, enabling us to examine the influence of the Bonnor-Melvin-$\Lambda$ model parameters through an analytical expression for the eigenfunctions. Here, we consider two scenarios for the vector potential: a linear and quadratic vector potentials. The problem is mapped into a Schrödinger-like equation for the first component of the DKP spinor with a Coulomb-like potential (linear vector potential) and harmonic oscillator (quadratic vector potential). The remaining components are expressed in terms of the one in a simple way. For the linear vector potential, the equation of motion, phase shift and scattering $S$-matrix are calculated from a Whittaker differential equation via partial wave analysis. The bound-state solutions are obtained from the poles of the $S$-matrix and the restriction on the potential parameters are discussed in detail. For the quadratic vector potential, the equation of motion and bound-state solutions are calculated from a confluent hypergeometric differential equation. In both scenarios, the first component of the DKP spinor is expressed in terms of the generalized Laguerre polynomial and the energy spectrum is composed of particle and antiparticle energies, and it is symmetrical about $E=0$.

This work is organized as follows. In section \ref{sec:1}, we give a brief review on DKP equation in a curved space-time. In section \ref{sec:eidkp}, we analyse the minimal coupling, the condition on the beta matrices which lead to a conserved current in a curved space-time, and the normalization condition for the DKP formalism. In the section \ref{sec:csb}, we focus in the DKP equation with minimal coupling in a Bonnor-Melvin-$\Lambda$ universe. Adopting the limit $\Lambda\ll 1$, we discuss in detail the problem considering two scenarios for the vector potential: linear potential (section \ref{sec:sub:li}) and quadratic potential (section \ref{sec:sub:qi}). Finally, in section \ref{sec:fr} we present our conclusions.                        

\section{Review of Duffin-Kemmer-Petiau equation in a curved space-time}
\label{sec:1}

The Duffin-Kemmer-Petiau (DKP) equation for a free boson in curved space-time is given by \cite{GRG34:491:2002,GRG34:1941:2002,EPJC75:287:2015,EPJC76:61:2016} ($\hbar =c=1$)%
\begin{equation}\label{dkp}
\left( i\beta ^{\mu }\nabla_{\mu}-M\right) \Psi =0
\end{equation}%
\noindent where the covariant derivative
\begin{equation}\label{der_cov}
\nabla_{\mu}=\partial _{\mu }-\Gamma_{\mu}\,.
\end{equation}
\noindent We restrict our study to the torsion-zero and the affine connection is defined by
\begin{equation}\label{affine}
\Gamma_{\mu}=\frac{1}{2}\,\omega_{\mu\bar{a}\bar{b}}[\beta^{\bar{a}},\beta^{\bar{b}}]\,.
\end{equation}
\noindent The curved-space beta matrices are
\begin{equation}\label{beta_curved}
\beta ^{\mu }=e^{\mu}\,_{\bar{a}}\,\beta^{\bar{a}}
\end{equation}
\noindent and satisfy the algebra%
\begin{equation}\label{betaalge}
\beta ^{\mu }\beta ^{\nu }\beta ^{\lambda }+\beta ^{\lambda }\beta
^{\nu }\beta ^{\mu }=g^{\mu \nu }\beta ^{\lambda }+g^{\lambda \nu
}\beta ^{\mu }\,.
\end{equation}%
\noindent where $g^{\mu \nu }$ is the metric tensor. The algebra represented by (\ref{betaalge}) gives rise to a set of 126 independent matrices, which can be classified into irreducible representations. These representations include a trivial representation, a five-dimensional representation that describes spin-zero particles (scalar sector), and a ten-dimensional representation associated to spin-one particles (vector sector). For the purposes of this discussion, we concentrate our attention on the spin-$0$ sector of the DKP theory.

The \textit{tetrads} $e_{\mu}\,^{\bar{a}}(x)$ satisfy the relations%
\begin{equation}\label{tetr1}
\eta^{\bar{a}\bar{b}}=e_{\mu}\,^{\bar{a}}\,e_{\nu}\,^{\bar{b}}\,g^{\mu\nu}
\end{equation}%
\begin{equation}\label{tetr2}
g_{\mu\nu}=e_{\mu}\,^{\bar{a}}\,e_{\nu}\,^{\bar{b}}\,\eta_{\bar{a}\bar{b}}
\end{equation}%
\noindent and
\begin{equation}\label{tetr3}
e_{\mu}\,^{\bar{a}}\,e^{\mu}\,_{\bar{b}}=\delta^{\bar{a}}_{\bar{b}}
\end{equation}%
\noindent the Latin indexes being raised and lowered by the Min\-kowski metric tensor $\eta^{\bar{a}\bar{b}}$ with signature $(-,+,+,+)$ and the Greek ones by the metric tensor $g^{\mu\nu}$.

The spin connection $\omega_{\mu\bar{a}\bar{b}}$ is given by
\begin{equation}\label{con}
\omega_{\mu}\,^{\bar{a}\bar{b}}=e_{\alpha}\,^{\bar{a}}\,e^{\nu \bar{b}}\,\Gamma_{\mu\nu}^{\alpha}
-e^{\nu \bar{b}}\partial_{\mu}e_{\nu}\,^{\bar{a}}
\end{equation}%
\noindent with $\omega_{\mu}\,^{\bar{a}\bar{b}}=-\omega_{\mu}\,^{\bar{b}\bar{a}}$ and $\Gamma_{\mu\nu}^{\alpha}$ are the Christoffel symbols given by
\begin{equation}\label{symc}
\Gamma_{\mu\nu}^{\alpha}=\frac{g^{\alpha\beta}}{2}\left( \partial_{\mu}g_{\beta\nu}+
\partial_{\nu}g_{\beta\mu}-\partial_{\beta}g_{\mu\nu} \right).
\end{equation}%
\noindent More detailed discussions on the DKP formalism in a curved space-time can be found in Ref.~\cite{EPJC75:287:2015}.
 
\section{Interactions in the DKP equation} 
\label{sec:eidkp}

The DKP equation for a charged scalar boson in curved space-time is given by
\begin{equation}
\left[ i\beta ^{\mu }\left(\nabla _{\mu }+iqA_{\mu}\right)-M\right] \Psi =0\,.  \label{dkpm1}
\end{equation}
\noindent As it is shown in Ref.~\cite{EPJC75:287:2015}, the conservation law for $J^{\mu}$ with interaction is given by \begin{equation}\label{corr}
\nabla_{\mu}J^{\mu}=\frac{1}{2}\bar{\Psi}\left(\nabla_{\mu}\beta^{\mu}\right)\Psi
\end{equation}
\noindent where $J^{\mu}=\frac{1}{2}\bar{\Psi}\beta^{\mu}\Psi$. It is worthwhile to mention that the interactions $\beta^{\mu}A_{\mu}$ is Hermitian with respect to $\eta^{0}$, with $\eta ^{0}=2\beta ^{0}\beta ^{0}-1$ (the matrices $\beta^{\mu}$ are Hermitian with respect to $\eta^{0}$, $(\eta^{0}\beta^{\mu})^{\dag}=\eta^{0}\beta^{\mu}$). Furthermore, if
\begin{equation}\label{cj0}
    \nabla_{\mu}\beta^{\mu}=0
\end{equation}
\noindent then four-current will be conserved \cite{EPJC75:287:2015,EPJC76:61:2016}. The condition (\ref{cj0}) is the purely geometrical assertion that the curved-space beta matrices are covariantly constant.

The normalization condition $\int{d\tau J^{0}}=\pm 1$ can be expressed as
\begin{equation}\label{normali1}
\int{d\tau \bar{\Psi}\beta^{0}\Psi}=\pm 2\,,
\end{equation}
\noindent where the plus (minus) sign must be used for a positive (negative) charge. Furthermore, the expectation value of any observable $\mathcal{O}$ can be given by
\begin{equation}\label{veo}
\langle \mathcal{O} \rangle=\frac{\int d\tau \bar{\Psi}\beta^{0}\mathcal{O}\Psi}{\int d\tau \bar{\Psi}\beta^{0}\Psi}\,,
\end{equation}
\noindent where $\beta^{0}\mathcal{O}$ should be Hermitian with respect to $\eta^{0}$, $\left[\eta^{0}\left(\beta^{0}\mathcal{O}\right)\right]^{\dagger}=\eta^{0}\left(\beta^{0}\mathcal{O}\right)$, in order to provide real eigenvalues \cite{PRA90:022101:2014}.

\section{DKP equation in a Bonnor-Melvin-$\Lambda$ universe}
\label{sec:csb}

The Bonnor-Melvin-$\Lambda$ solution of the Einstein-Maxwell equations to the case of a nonvanishing cosmological constant is described by the line element \cite{PRD99:044058:2019}
\begin{equation}\label{metric}
ds^{2}=-dt^{2}+dr^{2}+\sigma^{2}\sin^{2}\left(\sqrt{2\Lambda}r\right)d\varphi^{2}+dz^{2}\,,
\end{equation}
\noindent in cylindrical coordinates $(t,r,\varphi,z)$, where $-\infty<z<+\infty$, $r\geq 0$ and $0\leq\varphi\leq2\pi$. The parameter $\sigma$ is a integration constant and $\Lambda$ is the cosmological constant. The Ricci scalar curvature corresponding to this metric is $R= 4\Lambda$. This demonstrates that the spacetime described by the metric (\ref{metric}) is not asymptotically flat; instead, it displays a uniform curvature across the universe.

On the other hand, this line element reflects the curvature of spacetime determined by the following periodic magnetic field aligned with the axis of symmetry
\begin{equation}
H(r)=\sqrt{\Lambda}\sigma\sin\left(\sqrt{2\Lambda}r\right)   .
\end{equation}
The basis tetrad $e^{\mu}\,_{\bar{a}}$ from the line element (\ref{metric}) is chosen to be
\begin{equation}\label{tetra}
e^{\mu}\,_{\bar{a}}=%
\begin{pmatrix}
1 & 0 & 0 &  0\\
0 & 1 & 0 & 0\\
0 & 0 & \frac{1}{\sigma\sin\left(\sqrt{2\Lambda}r\right)} & 0\\
0 & 0 & 0 & 1
\end{pmatrix}%
\,.
\end{equation}%
\noindent For the specific basis tetrad (\ref{tetra}) the curved-space beta matrices read
\begin{eqnarray}
\beta^{0} &=& \beta^{\bar{0}}\,,  \label{betat} \\
\beta^{r} &=& \beta^{\bar{1}}\,,   \label{betar} \\
\beta^{\varphi} &=& \frac{1}{\sigma\sin\left(\sqrt{2\Lambda}r\right)}\beta^{\bar{2}}\,,  \label{betaphi}\\
\beta^{z}&=& \beta^{\bar{3}}\,, \label{betaz}
\end{eqnarray}%
\noindent and the spin connection is given by
\begin{equation}\label{gphi}
\Gamma_{\varphi}=-\sqrt{2\Lambda}\sigma\cos\left(\sqrt{2\Lambda}r\right)\left[\beta^{\bar{1}},\beta^{\bar{2}}\right]\,.
\end{equation}
\noindent Thereby, the covariant derivative gets
\begin{eqnarray}
\nabla_{0} &=& \partial_{0}\,,\label{dt}\\
\nabla_{r} &=& \partial_{r}\,,\label{dr}\\
\nabla_{\varphi} &=& \partial_{\varphi}+\sqrt{2\Lambda}\sigma\cos\left(\sqrt{2\Lambda}r\right)\left[\beta^{\bar{1}},\beta^{\bar{2}}\right]\,,\label{dphi}\\
\nabla_{z} &=& \partial_{z}
\end{eqnarray}
\noindent At this stage, note that the condition (\ref{cj0}) is satisfied for the curved-space beta matrices (\ref{betat}), (\ref{betar}), (\ref{betaphi}) and (\ref{betaz}) and therefore the current is conserved for this background. 

Following the expressions presented above, we now study the dynamics of a charged scalar boson in a Bonnor-Melvin-$\Lambda$ universe in the presence of an additional magnetic field. To accomplish this, we consider a four-vector potential, $A_{\mu}$, in curved space associated with this external field. This potential is defined as $A_{\mu} = e_{\mu}{}^{\bar{a}} A_{\bar{a}}$, where $A_{\bar{a}}$ represents the four-vector potential in flat space-time \cite{oliveira2020noninertial, huamani2022aharonov, soares2023effects}. In this case, we consider  $A_{\bar{a}}$ in the form
\begin{equation}
A_{\bar{a}}=(0,0,A_{\bar{2}},0)=(0,0,-f(r),0), 
\end{equation}
where $f(r)$ is the functional form of the four-vector potential in the flat space-time. The choice of this structure for $A_{\bar{a}}$ was driven by the goal of preserving the axial symmetry of the Bonnor-Melvin-$\Lambda$ universe. Furthermore, this configuration facilitates the derivation of analytical solutions under the approximation described in this work. We will discuss this approach in more detail later.  

By using the inverse tetrad $e_{\varphi}\,^{\bar{2}}=\sigma \sin\left(\sqrt{2 \Lambda}r \right)$, 
the four-vector potential in the Bonnor-Melvin-$\Lambda$ universe is defined as
\begin{equation}\label{Aphi}
A_{\varphi}=\sigma\sin\left(\sqrt{2\Lambda}r\right)f(r)\,.
\end{equation}

\noindent As the interaction is time-independent, we can express $\Psi(\vec{r},t)=\Phi(\vec{r})\mathrm{exp}\left(-iEt\right)$ in equation (\ref{dkpm1}), where $E$ represents the energy of the scalar boson. Consequently, the time-independent DKP equation transforms into
\begin{equation}\label{dkpm2}
\left[ \beta^{\bar{0}}E+i\beta^{\bar{1}}\partial_{r}+i\beta^{\varphi}
\nabla_{\varphi}+\beta^{\varphi}A_{\varphi}+i\beta^{\bar{3}}\partial_{z}-M\right]\Phi=0\,,
\end{equation}
\noindent where $\beta^{\varphi}$, $\nabla_{\varphi}$ and $A_{\varphi}$ are given by (\ref{betaphi}), (\ref{dphi}) and (\ref{Aphi}), respectively.  We now adopt the standard representation for the beta matrices as described in \cite{JPG19:87:1993}. With the five-component spinor $\Phi^T=(\Phi_{1},\ldots,\Phi_{5})$, the time-independent DKP equation (\ref{dkpm2}) becomes
\begin{eqnarray}
E\Phi_{2}&-&M\Phi_{1}-i\bar{\partial}_{r}\Phi_{3}-i\bar{\partial}_{\varphi}\Phi_{4}-i\partial_{z}\Phi_{5}=0\,,\label{phi1}\\
\Phi_{2} &=& \frac{E}{M}\Phi_{1} \,,  \label{phi2} \\
\Phi_{3} &=& \frac{i}{M}\partial_{r}\Phi_{1} \,,   \label{phi3} \\
\Phi_{4} &=& \frac{i}{M}\bar{\partial}_{\varphi}\Phi_{1}  \,,  \label{phi4}\\
\Phi_{5}&=& \frac{i}{M}\partial_{z}\Phi_{1} \,, \label{phi5}
\end{eqnarray}%
\noindent where
\begin{eqnarray}
 \bar{\partial}_{r} &=& \partial_{r}+\frac{\sqrt{2\Lambda}}{\tan\left(\sqrt{2\Lambda}r\right)}\,, \\
 \bar{\partial}_{\varphi} &=& \frac{\partial_{\varphi}}{\sigma\sin\left(\sqrt{2\Lambda}r\right)}-iqf(r)\,.
\end{eqnarray}
\noindent Nonetheless,
\begin{equation}\label{j0}
J^{0}=\mathrm{Re}\left(\Phi_{2}^{\ast}\Phi_{1}\right)=\frac{E}{M}\vert \Phi_{1}\vert^{2}\,.
\end{equation}
\noindent By combining these outcomes, we establish an equation of motion governing the first component of the DKP spinor
\begin{equation}\label{mg82}
\begin{split}
    &\left[\partial^2_r+ \frac{\sqrt{2\Lambda}}{\tan \sqrt{2\Lambda}r}\partial_r+\frac{1}{\sigma^{2}\sin^{2}\sqrt{2\Lambda}r}\partial^{2}_{\varphi}+\partial^2_z\right.\\
     &\left.-q^{2}f^{2}(r)-\frac{2iqf(r)}{\sigma\sin \sqrt{2\Lambda}r}\partial_{\varphi}+E^{2}-M^{2}\right]\Phi_{1}=0\,. 
\end{split}
\end{equation}
Given the complexity of equation (\ref{mg82}), we adopted an approach that allows us to analyze the impact of the Bonnor-Melvin-$\Lambda$ model parameters on scalar boson dynamics using analytical solutions for $\Phi_1$. We explain our reasoning as follows.

As mentioned earlier, our investigation focused on examining the quantum effects associated with gravitation from the perspective of relativistic quantum mechanics in curved spacetimes. Several studies have been instrumental in exploring these effects, including the use of the Dirac \cite{dariescu2021dirac} and Klein-Gordon \cite{vieira2014exact} equations in Kerr-Newman space, as well as the Dirac equation in Robertson-Walker spacetime \cite{barut1987exact}, among others. Nonetheless, one aspect that has not received enough attention in this context is related to distance scales. In the context of the Bonnor-Melvin-$\Lambda$ universe, it becomes particularly interesting to examine specific spatial regions where the order of magnitude of spatial coordinate $r$ is inversely proportional to the order of magnitude of $\sqrt{\Lambda}$. This proportionality plays a crucial role in preserving the effects of the cosmological constant $\Lambda$ and ensuring the periodic behavior of both the metric and the magnetic field, due to the presence of the term $\sin(\sqrt{2\Lambda}r)$. This approach was utilized in the works of Žofka in the context of general relativity, as detailed in references \cite{PRD99:044058:2019}, \cite{vesely2021cylindrical}, and \cite{vesely2022exact}. In this manner, the cosmological constant $\Lambda$ effectively controls the order of magnitude of cosmological distances, manifesting its influence at such scales.

Nevertheless, our work focuses on a quantum scenario, rather than a cosmological one. That is, the spatial coordinate $r$ is at a microscopic scale. Consequently, the cosmological constant is no longer capable of controlling the order of magnitude of the argument of the sine, making the product $\sqrt{2\Lambda}r$ very small, or, equivalently, $\Lambda\ll 1$. The plausibility of a small cosmological constant has been considered in various scenarios at quantum scale, including non-supersymmetric theories with a light dilaton \cite{EPJC74:2790:2014}, 5D brane world models \cite{JHEP2017:71:2017}, and supersymmetric string theory \cite{GRG40:607:2008,PRL128:011602:2022}, among others.

It is noteworthy that this condition naturally supports the approximation $\sin(\sqrt{2\Lambda}r) \approx \sqrt{2\Lambda}r$, allowing the metric (\ref{metric}) to be reexpressed as
\begin{equation}
    ds^2=-dt^2+dr^2+2 \Lambda \sigma^2 r^2 d\varphi^2 +d z^2 \label{metric2},
\end{equation}
where the Ricci scalar curvature is now given by $R = \delta(r)$. This implies that the curvature in this metric is null everywhere except at the point where the source is located $(r=0)$, resembling the spacetime generated by a cosmic string \cite{EPJC75:287:2015}. From the latter, one can infer that the metric (\ref{metric2}) describes a spacetime with a conical singularity generated by the term $2 \Lambda \sigma^2$ at $r=0$ \cite{sokolov1977structure}. Furthermore, the line element (\ref{metric2}) produces a magnetic field aligned with the axis of symmetry given by
\begin{equation}
H(r)=\sqrt{2}\Lambda\sigma r\,,
\end{equation}
\noindent and the four-vector potential in the Bonnor-Melvin-$\Lambda$ universe (\ref{Aphi}) reduces to
\begin{equation}
A_{\varphi}=\sqrt{2\Lambda}\sigma rf(r)\,.
\end{equation}

Therefore, in this approximation, the equation (\ref{mg82}) becomes 
\begin{equation}\label{mg83}
\begin{split}
    &\left[\partial^2_r+ \frac{1}{r}\partial_r+\frac{1}{H_{0}^{2}r^{2}}\partial^{2}_{\varphi}+\partial^2_z-q^{2}f^{2}(r)\right.\\
     &\left.-\frac{2iqf(r)}{H_{0}r}\partial_{\varphi}+E^{2}-M^{2}\right]\Phi_{1}=0\,, 
\end{split}
\end{equation}
\noindent where $H_{0}=\sqrt{2\Lambda}\sigma$. At this point, we can leverage the invariance under boosts along the $z$-direction and proceed with the standard decomposition:
\begin{equation}\label{ansatz}
\Phi_{1}(r,\varphi,z)=\frac{\phi_{1}(r)}{\sqrt{r}}e^{im\varphi+ik_{z}z}
\end{equation}
\noindent with $m\in\mathbb{Z}$. Inserting this into Eq. (\ref{mg83}), we obtain
\begin{equation}\label{ed2o}
\begin{split}
&\left[\frac{d^2}{dr^2}-\frac{\left(m_{\Lambda}^2-\frac{1}{4}\right)}{r^2}-q^{2}f^{2}(r)\right.\\
&\left.+\frac{2qm_{\Lambda}f(r)}{r}+E^{2}-M^{2}-k_{z}^{2}\right]\phi_{1}=0\,,
\end{split}
\end{equation}
\noindent where $m_{\Lambda}=\frac{m}{H_{0}}$. Now, we can apply the framework developed above to solve the DKP equation in this background, utilizing specific forms for $f(r)$. The choice of $f(r)$ is justified by the magnetic field produced by the four-vector potential, which is expected to be uniform as indicated in reference \cite{vallee2004cosmic}.

\subsection{Linear interaction}
\label{sec:sub:li}
Considering $f(r)=\frac{\alpha_{1}}{2}$, where $\alpha_{1}$ is a constant, this particular form of $f(r)$ leads to $A_{\varphi}=\frac{\alpha_{1} H_{0}r}{2}$ (linear interaction), which furnish a magnetic field $\vec{B}=\frac{\alpha_1}{2r}\hat{z}$. In this scenario, the equation (\ref{ed2o}) becomes
\begin{equation}\label{ed2o2}
\left[-\frac{d^2}{dr^2}-\frac{\delta}{r}+\frac{\left(m_{\Lambda}^2-\frac{1}{4}\right)}{r^2}-\kappa^2\right]\phi_{1}=0\,,
\end{equation}
\noindent where $\delta=qm_{\Lambda}\alpha_{1}$ and 
\begin{equation}
\kappa=\sqrt{E^2-M^2-k_{z}^{2}-\frac{q^{2}\alpha_{1}^{2}}{4}}\,.
\end{equation}
\noindent Equation (\ref{ed2o2}) characterizes the relativistic motion of charged scalar bosons that interact with a linear vector potential within the framework of a Bonnor-Melvin-$\Lambda$ universe. With $\phi_{1}(0)=0$ and $\int_{0}^{\infty}\vert \phi_{1}\vert^{2}dr<\infty$, this equation exactly corresponds to the time-independent radial Schrödinger equation for a Coulomb-like potential in two dimensions. The potential exhibits a well structure when $\delta>0$, which implies that $m_{\Lambda}\alpha_{1}>0$. Additionally, bound states are expected for $\kappa=i\vert\kappa\vert$, i.e. $-\varepsilon<E<+\varepsilon$, with
\begin{equation}\label{eplusminus}
\varepsilon=\sqrt{M^{2}+k_{z}^{2}+\frac{q^{2}\alpha_{1}^{2}}{4}}\,.
\end{equation}   
\noindent Therefore, we can conclude that bound-state solutions might occur only for $m_{\Lambda}\alpha_{1}>0$, corresponding to energies in the interval $-\varepsilon<E<+\varepsilon$. On the other hand, the solution $\phi_{1}(r)$ has asymptotic behavior $\mathrm{e}^{\pm i\kappa r}$, so we can expect that scattering states only occur if $\kappa\in \mathbb{R}$, i.e. $\vert E\vert>\varepsilon$.

%After carrying out a qualitative analysis of the effective potential, we are able to make a quantitative approach of our problem, by examining both scattering and bound states.   
After conducting a qualitative analysis of the effective potential, we will be able to develop a quantitative approach to our problem by examining the bound-state solutions, which are obtained from the poles of the S-matrix, and the restrictions on the potential parameters of the system. This will be discussed in detail in the following section.

\subsubsection{Scattering states and the $S$-matrix}

\begin{figure}[H]
\begin{subfigure}{0.39\textwidth}
\centering
\includegraphics[width=0.93\linewidth,angle=0]{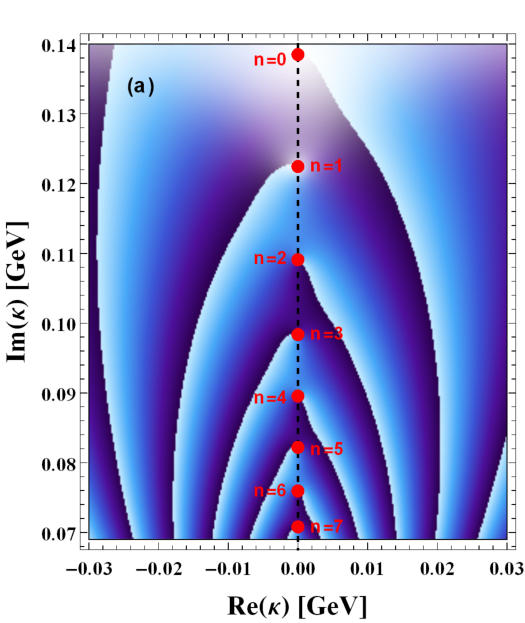}
\label{fig:EvsOmega_l0}
\end{subfigure}
\begin{subfigure}{0.39\textwidth}
\centering
\includegraphics[width=0.93\linewidth,angle=0]{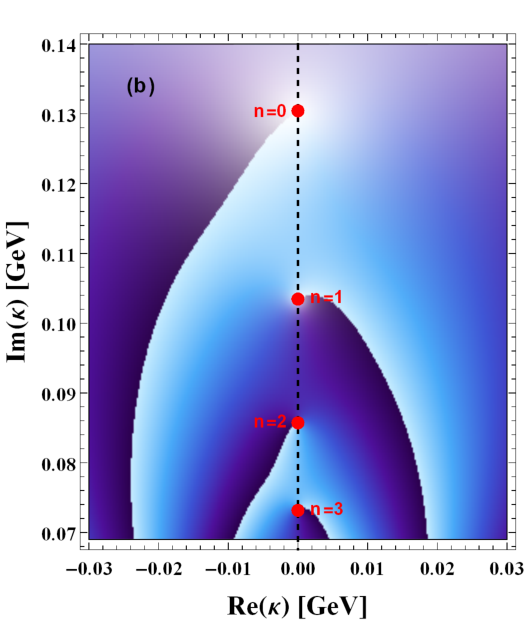}
\label{fig:EvsOmega_l1}
\end{subfigure}
\begin{subfigure}{0.39\textwidth}
\centering
\includegraphics[width=0.93\linewidth,angle=0]{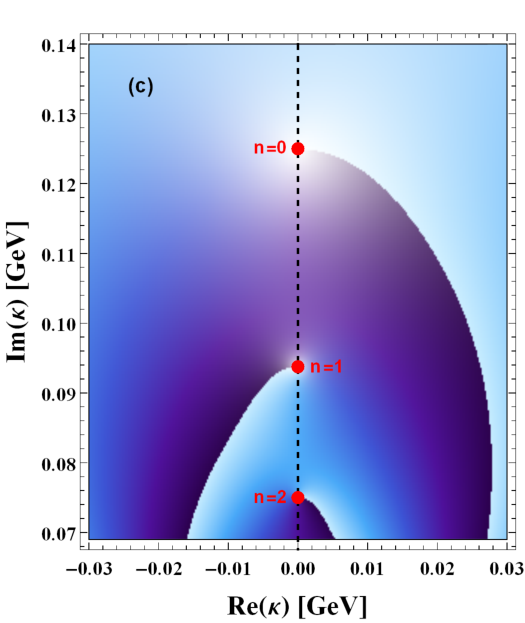}
\label{fig:EvsOmega_l2}
\end{subfigure}
\caption{\label{figS} Plot of the $S$-matrix as a function of the real and imaginary part of $\kappa$ for $\Lambda=0.011$ (panel a), $\Lambda=0.045$ (panel b) and $\Lambda=0.080$ (panel c) (expressed in natural units).}
\label{fig:EvsOmega}
\end{figure}

Using the abbreviation
\begin{equation}
\eta=\frac{\delta}{2\kappa} ,
\end{equation} 
and the change $\rho=-2i\kappa r$, the equation (\ref{ed2o2}) becomes
 \begin{equation}
    \frac{d\phi_{1}}{d\rho^2}+\left(-\frac{1}{4}+\frac{i\eta}{\rho}+\frac{\frac{1}{4}-m_{\Lambda}^2 }{\rho^2} \right)\phi_{1} =0 . \label{eq5}
\end{equation}
\noindent This second-order differential equation is the called Whittaker
equation. Owing to the condition $\phi_{1}(0)=0$, only the regular solution at the origin is allowed. In this case, the solution is given by
\begin{equation}
\phi_{1}\propto \rho^{\vert m_{\Lambda}\vert+1/2} e^{-\rho/2} M\left(a,b;\rho\right)\,, \label{phi1sol}
\end{equation}
\noindent where 
\begin{eqnarray}
a &=& \vert m_{\Lambda}\vert-i\eta+\frac{1}{2}\,,\\
b &=& 1+2\vert m_{\Lambda}\vert\,, 
\end{eqnarray}
\noindent and $M(a,b; \rho)$ is the Kummer's function, whose asymptotic behavior for large $|\rho|$ with a purely imaginary $\rho=-i|\bar{\rho}|$, where $|\bar{\rho}|=2\kappa r$ is given by \cite{JAMES1991} 
\begin{equation}\label{com_as}
\begin{split}
M(a,b;\rho) & \simeq \frac{\Gamma(b)}{\Gamma(a-b)}e^{-\frac{i\pi a}{2}}|\bar{\rho}|^{-a}\\
&+\frac{\Gamma(b)}{\Gamma(a)}e^{-i[|\bar{\rho}|-\frac{\pi(b-a)}{2}]}|\bar{\rho}|^{a-b}.
\end{split}
\end{equation}
\noindent Using this last result, we can show that for $\vert\rho\vert\gg 1$ and $\kappa \in \mathbb{R}$, the asymptotic behavior of (\ref{phi1sol}) is given by \cite{AP146:1:1983}
\begin{equation}
\phi(r)\simeq \cos \left(\kappa r-\frac{\vert m\vert\pi}{2}-\frac{\pi}{4}+\delta_m\right),
\end{equation}
\noindent where the relativistic phase shift $\delta_m = \delta_m(\eta)$ is given by
\begin{equation}\label{phase_shift}
\delta_m=\frac{\pi}{2}(\vert m\vert-\vert m_{\Lambda}\vert)+ \text{arg} \Gamma \left(\frac{1}{2}+\vert m_{\Lambda}\vert-i\eta\right)\,.
\end{equation}
\noindent From this last expression, we can express the scattering $S$-matrix as
\begin{equation}\label{s_matrix}
S_{m}=\mathrm{e}^{2i\delta_{m}}=\mathrm{e}^{i\pi(\vert m\vert-\vert m_{\Lambda}\vert)}\frac{\Gamma\left(1/2+\vert m_{\Lambda}\vert-i\eta\right)}{\Gamma\left(1/2+\vert m_{\Lambda}\vert+i\eta\right)}\,.
\end{equation} 
\noindent The distribution of points in the complex plane of the $S$-matrix is shown in figure \ref{figS}, with the following parameters \cite{PRD97:076008:2018}: with $m=1$, $q=1$, $M=0.14$\,GeV, $k_{z}=0.14$\,GeV, $\alpha_1 = 0.30$\,GeV, $\sigma=1$ and $\Lambda=0.011$ (panel a), $\Lambda=0.045$ (panel b) and $\Lambda=0.080$ (panel c) (expressed in natural units). 

The red points in Fig. \ref{figS} represent the poles of the $S$-matrix, which are infinite in number and located on the imaginary axis of $\kappa$ (where $\mathrm{Re}(\kappa) = 0$). These poles occur only for non-negative integer values of the argument of the gamma function in the numerator of equation (\ref{s_matrix}). In each panel, the $\kappa$ values associated with an $S$-matrix pole exhibit a maximum at the zeroth-order pole, which is given by $\kappa_{max} = i\delta/(1+2\vert m_{\Lambda}\vert)$. Note that the spacing between the poles decreases progressively as higher-order poles are observed. Notably, as $\Lambda$ increases ($\vert m_{\Lambda}\vert$ decreases), these poles tend to shift towards $\mathrm{Im}(\kappa) = 0$, leading to an increase in the concentration of poles in the regions of the higher-order poles. For this reason, Fig. \ref{figS} specifically selects an interval of 
\noindent $\mathrm{Im}(\kappa)$ from $0.07$ to $0.14$ on the gigascale for  each panel, allowing for a clear visualization of this redistribution of the poles as the cosmological constant $\Lambda$ varies. A high concentration of poles is associated with the so-called accumulation point of states, which is a distinctive feature in systems with Coulomb-like interactions that exhibit an infinite number of states below a certain energy. This observation is crucial in the next section, where we establish that the presence of these poles in the $S$-Matrix leads to the quantization condition for bound-state solutions, as discussed in \cite{castro2017relativistic,neto2020scalar}.

\subsubsection{Bound states}

The energies of the bound-state solutions can be obtained from poles of the $S$-matrix when one considers $\kappa$ imaginary. Then, if $\kappa=i\vert \kappa\vert$, the $S$-matrix becomes infinite when $1/2+\vert m_{\Lambda}\vert-i\eta=-n$, where $n=0,1,2,\ldots$, owing to the poles of the gamma function in the numerator of (\ref{s_matrix}). As previously indicated, bound-state solutions are possible only for $m_{\Lambda}\alpha_{1}>0$, corresponding to energies in the interval $-\varepsilon<E<+\varepsilon$ and the spectrum is given by
\begin{equation}\label{ener1}
E=\pm\sqrt{M^{2}+k_{z}^{2}+\frac{q^{2}\alpha_{1}^{2}}{4}-\frac{q^{2}\alpha_{1}^{2}m_{\Lambda}^{2}}{4\mu^{2}}}\,,
\end{equation}  
\noindent with $\mu=n+1/2+\vert m_{\Lambda}\vert$. Note that the energy spectrum, comprising both particle energy (positive) and antiparticle energy (negative), is symmetrical about $E=0$, indicating that this scenario does not distinguish between particles and antiparticles. Furthermore, the energy spectrum (\ref{ener1}) features an infinite number of bound states, with each successive pair of energies being closer together than the previous pair. This means that sufficiently high pairs of energy levels approach spacing near zero. In this region of the spectrum, the effects of energy discretization diminish, thereby bringing the system closer to the so-called accumulation point of states as $E \rightarrow \varepsilon$, occurring when $n \rightarrow \infty$. This accumulation point establishes the threshold between the spectrum of bound states and scattering states, which is a characteristic of systems with Coulomb-like interactions. In this scenario, the cosmological constant acts as a control parameter, accelerating the accumulation of states as $\Lambda$ increases ($m_\Lambda$ decreases), and vice versa. This finding is consistent with the results shown in Fig. \ref{fig:EvsOmega}, where the concentration of poles increases as $\sqrt{2\Lambda}\sigma$ increases.

\begin{figure}[t]
\includegraphics[width=0.93\linewidth,angle=0]{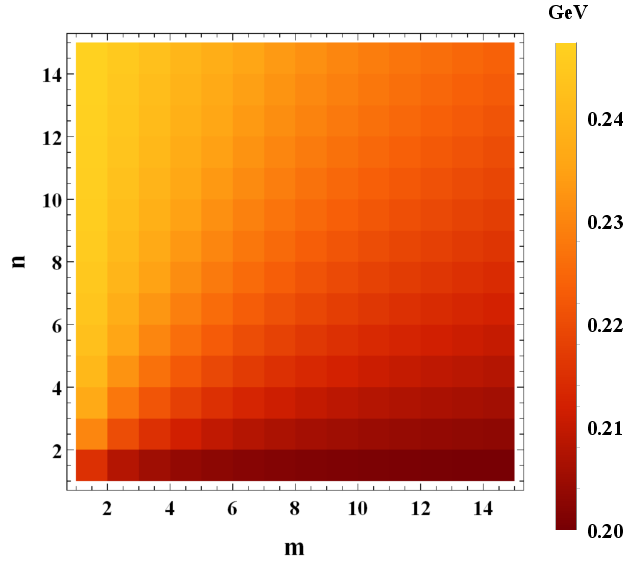}
\caption{\label{fige1} Plot of the energy distribution $\vert E\vert$ as a function of the quantum numbers $n$ and $m$, with $q=1$, $M=0.14$\,GeV, $k_{z}=0.14$\,GeV, $\sigma=1$, $\Lambda=0.80$ and $\alpha_1 = 0.30$\,GeV (expressed in natural units).}
\end{figure}

On the other hand, in Fig. \ref{fige1} we illustrate the energy distribution $\vert E\vert$ for bound states as a function of the quantum numbers $n$ and $m$, with $q=1$, $M=0.14$\,GeV, $k_{z}=0.14$\,GeV, $\sigma=1$, $\Lambda=0.80$ and $\alpha_1 = 0.30$\,GeV (expressed in natural units). At this point, we would like to highlight three specific situations. In the case, when $n\gg m$, the energies of bound states converge to approximately $0.24$\, GeV in regions near the vertical axis. This energy value corresponds to the upper threshold defined in (\ref{eplusminus}). Conversely, when $n/m\approx 1$ (along the diagonal of the figure), we can observe that the energies cluster around $0.22$ \, GeV, remaining fixed regardless of the values of $n$ and $m$.

Finally, in the situation when $n\ll m$, the energies reach their lowest value of approximately $0.20$ in regions near the horizontal axis. The solution for all $r$ can be written as
\begin{equation}\label{solu}
\phi_{1}(\rho)=N_{n} r^{\vert m_{\Lambda}\vert+1/2}\mathrm{e}^{-\vert\kappa\vert r}L^{(2\vert m_{\Lambda}\vert)}_{n}(2\vert \kappa\vert r)\,,
\end{equation}
\noindent due to $M(-n,b;\rho)$ with $b>0$ is proportional to the generalized Laguerre polynomial $L^{(b-1)}_{n}(\rho)$ \cite{ABRAMOWITZ1965}, a polynomial of degree $n$ with $n$ distinct positive zeros in the range $[0,\infty)$. The normalization constant $N_{n}$ in (\ref{solu}) is obtained by means of the normalization condition. The charge density $J^{0}$ (\ref{j0}) implies that $\phi_{1}$ must be normalized as
\begin{equation}\label{norma_c}
\frac{\vert E\vert}{M}\int dr \vert\phi_{1}\vert^{2}=1\,.
\end{equation}
\noindent In this way, the normalization constant can be expressed as
\begin{equation}\label{cn1}
N_{n}=\sqrt{\frac{M(2\vert\kappa\vert)^{2\vert m_{\Lambda}\vert+2}n!}{\vert E\vert(2n+2\vert m_{\Lambda}\vert+1)[(n+2\vert m_{\Lambda}\vert)!]^{3}}}\,,
\end{equation} 
\noindent with $\vert E\vert\neq 0$. 
%\newpage
\subsection{Quadratic interaction}
\label{sec:sub:qi}

Considering $f(r)=\frac{\alpha_{2}r}{2}$, where $\alpha_{2}$ is a constant, this particular form of $f(r)$ leads to $A_{\varphi}=\alpha_{2}H_{0}r^{2}/2$ (quadratic interaction). The choice of the functional form of the four-vector potential was made based on the magnetic field produced by it, which is expected to be a constant magnetic field  $\vec{B}=\alpha_2\hat{z}$, as indicated in reference \cite{durrer2007cosmic}.  In this scenario, the equation (\ref{ed2o}) becomes
\begin{equation}\label{ed2o3}
\left[-\frac{d^2}{dr^2}+\lambda^{2}r^{2}+\frac{\left(m_{\Lambda}^2-\frac{1}{4}\right)}{r^2}-K^2\right]\phi_{1}=0\,,
\end{equation}
\noindent where $\lambda=q\vert\alpha_{2}\vert/2$ and 
\begin{equation}
K=\sqrt{E^2-M^2-k_{z}^{2}+qm_{\Lambda}\alpha_{2}}\,.
\end{equation}
\noindent Equation (\ref{ed2o3}) defines the relativistic motion of charged scalar bosons interacting with a quadratic vector potential in the context of a Bonnor-Melvin-$\Lambda$ universe. With $\phi_{1}(0)=0$ and $\int_{0}^{\infty}\vert \phi_{1}\vert^{2}dr<\infty$, the solution for (\ref{ed2o3}) with $K$ and $\lambda$ real exactly corresponds to the well-known solution of the time-independent radial Schrödinger equation for a harmonic oscillator in two dimensions. Note that the condition $K\in\mathbb{R}$ implies that $\vert E\vert>\epsilon$, where
\begin{equation}\label{epsi}
\epsilon=\sqrt{M^{2}+k_{z}^{2}-qm_{\Lambda}\alpha_{2}}\,,
\end{equation}
\noindent for $M^{2}+k_{z}^{2}-qm_{\Lambda}\alpha_{2}\geqslant 0$, and for all $E$ if $M^{2}+k_{z}^{2}-qm_{\Lambda}\alpha_{2}<0$. In this scenario, we only find bound-state solutions. 

\subsubsection{Bound states}

Considering the solution for all $r$ in the form 
\begin{equation}\label{ansatz2}
\phi_{1}(r)=r^{\vert m_{\Lambda}\vert+\frac{1}{2}}e^{-\lambda r^2/2}g(r)\,,
\end{equation} 
\noindent and, along with the definition of a new variable and parameters:
\begin{eqnarray}
 \rho &=& \lambda r^2\,, \label{newvar}\\
 a &=& \frac{1}{2}\left(\vert m_{\Lambda}\vert+1-\frac{K^2}{2\lambda}\right)\,,\label{a}\\
 b &=& \vert m_{\Lambda}\vert+1\,,\label{b}
\end{eqnarray}
\noindent we find that $g(\rho)$ satisfies a confluent hypergeometric equation \cite{ABRAMOWITZ1965},
\begin{equation}\label{hipergeome}
\rho\frac{d^2g}{d\rho^2}+\left(b-\rho\right)\frac{dg}{d\rho}-ag=0\,.
\end{equation}
\noindent Owing to the condition $\phi_{1}(0)=0$, the regular solution at the origin is given by
\begin{equation}\label{gs}
g(\rho)\propto M\left(a,b,\rho\right)\,.
\end{equation}
\noindent The asymptotic behavior of Kummer's function is dictated by \cite{ABRAMOWITZ1965}
\begin{equation}\label{asymp}
M\left(a,b,\rho\right)\simeq \frac{\Gamma(b)}{\Gamma(b-a)}e^{-i\pi a}\rho^{-a}+
\frac{\Gamma(b)}{\Gamma(a)}e^{\rho}\rho^{a-b}\,.
\end{equation}
\noindent It is true that the presence of term $e^{\rho}$ in the asymptotic behavior of $M(a,b,\rho)$ perverts the normalizability of $\phi_{1}(\rho)$. Nevertheless, this trouble can be surpassed by demanding $a=-n$, where $n$ is a nonnegative integer and $b\neq-\tilde{n}$, where $\tilde{n}$ is also a nonnegative integer. In fact, $M(-n,b,\rho)$ with $b>0$ is proportional to the generalized Laguerre polynomial $L_{n}^{(b-1)}(\rho)$, a polynomial of degree $n$ with $n$ distinct positive zeros in the range $[0,\infty)$. Therefore, the solution for all $r$ can be written as
\begin{equation}\label{solg}
\phi_{1}(r)=N_{n}r^{|m_{\Lambda}|+\frac{1}{2}}e^{-\lambda r^2/2}L^{(\vert m_{\Lambda}\vert)}_{n}(\lambda r^{2})\,,
\end{equation}
\noindent where $N_{n}$ is a normalization constant. The charge density $J^{0}$ (\ref{j0}) implies that
$\phi_{1}$ must be normalized as
\begin{equation}\label{norma3}
\frac{|E|}{M}\int_{0}^{\infty}dr|\phi_{1}|^{2}=1\,,
\end{equation}
\noindent so that, the normalization constant can be expressed as
\begin{equation}\label{norma4}
N_{n}=\sqrt{\frac{2M\lambda^{|m_{\Lambda}|+1}\Gamma\left(n+1\right)}{|E|\Gamma\left(|m_{\Lambda}|+n+1\right)}}\,,
\end{equation}
\noindent with $\vert E\vert\neq 0$. Furthermore, the requirement $a=-n$ (quantization condition) implies into
\begin{equation}\label{energyr}
E=\pm\sqrt{M^2+k_{z}^{2}+q\vert\alpha_{2}\vert\left(2n+\vert m_{\Lambda}\vert-m_{\Lambda}\mathrm{sgn}(\alpha_{2})+1\right)}\,,
\end{equation}
\noindent where $\mathrm{sgn}(x)$ is the sign function. From (\ref{energyr}), we can see that Landau levels (energy independent of $m$) might occur if only if $m_{\Lambda}\alpha_{2}>0$. Note that the energy spectrum 

\begin{figure}[h]
\includegraphics[width=0.93\linewidth,angle=0]{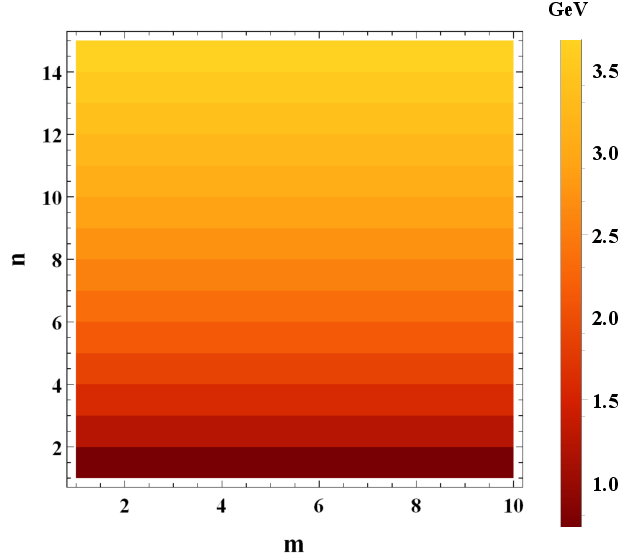}
\caption{\label{fige2} Plot of the energy distribution $\vert E\vert$ as a function of the quantum numbers $n$ and $m$, with $q=1$, $M=0.14$\,GeV, $k_{z}=0.14$\,GeV, $\sigma=1$, $\Lambda=0.80$ and $\alpha_2 = 0.50$\,GeV (expressed in natural units).}
\end{figure}  

\noindent is composed of particle energy (positive energy) and antiparticle energy (negative energy) and it is symmetrical about $E=0$. As in the previous scenario, we can conclude that this scenario does not distinguish particles from antiparticles. In Fig. \ref{fige2} displays the energy distribution $\vert E\vert$ for bound states as a function of the quantum numbers $n$ and $m$, with $q=1$, $M=0.14$\,GeV, $k_{z}=0.14$\,GeV, $\sigma=1$, $\Lambda=0.80$ and $\alpha_2 = 0.50$\,GeV (expressed in natural units). Here, the cosmological constant, specifically the product $\sqrt{2\Lambda}\sigma$, affects the minimum of the effective potential well, promoting the emergence of more energetic bound states when $\sqrt{2\Lambda}\sigma$ decreases, and vice versa. In this case, we observe the emergence of traditional Landau levels distributed in horizontal bars of varying intensities. Such a characteristic can be explained by the positivity of $\alpha_2$ (as per experimental data),  which implies that $m$ must also be

\noindent positive in order to satisfy the condition $m_{\Lambda}\alpha_{2}>0$. Furthermore, in contrast to the linear case, the absence of an upper energy threshold in this scenario is attributed to the fully confining nature of the effective potential (\ref{ed2o3}).

\section{Conclusions}
\label{sec:fr}

In conclusion, this study was motivated by previous research and aimed to explore the quantum dynamics of a charged scalar boson within the framework of the DKP formalism while immersed in a Bonnor-Melvin-$\Lambda$ universe. The investigation focused on the scalar sector of the DKP equation, yielding the equation of motion for a arbitrary vector potential. Due to the inherent complexity of the equation of motion, we were compelled to employ the limit $\Lambda\ll 1$ to obtain exact solutions. We considered two distinct scenarios for the vector potential: a linear potential and a quadratic potential. In both cases, the problem was transformed into a Schrödinger-like equation for the first component of the DKP spinor. This resulted in Coulomb-like and harmonic oscillator potentials for the linear and quadratic vector potentials, respectively. The remaining components were expressed in a straightforward manner in terms of the first one. For the linear vector potential, the bound-state solutions were determined by identifying the poles of the $S$-matrix. It was observed that the cosmological constant acts as a control parameter, accelerating the accumulation of states at higher energy levels as $\Lambda$ increases, and vice versa. This observation aligns with the findings presented in Fig. \ref{fig:EvsOmega}, which show an increase in the concentration of poles as $\sqrt{2\Lambda}\sigma$ increases.  Additionally, it was also observed that the parameters of the problem must satisfy the restriction  $m_\Lambda \alpha_1 >0$. In this context, we obtained a symmetric energy spectrum that includes both positive and negative energies, implying an indistinguishability between particles and antiparticles. On the other hand, for the quadratic vector potential, the equation of motion and bound-state solutions were derived from a confluent hypergeometric differential equation. In this scenario, it was observed that the product $\sqrt{2\Lambda}\sigma$ affects the minimum of the effective potential well, promoting the emergence of more energetic bound states when this term decreases, and vice versa. Furthermore, we note that the energy expression is similar to the spectrum shown in the Landau quantization, if only if the restriction on the problem parameters $m_{\Lambda}\alpha_{2}>0$ is satisfied. It is worth highlighting  that the energy expression for the bound states, despite its structure, also shows symmetry with respect to $E=0$, as in the previous scenario. Also, we found the normalized solutions for both scenarios expressed in terms of the generalized Laguerre polynomials. The findings in this article suggest that the cosmological constant plays a crucial role in the emergence of bound states and energy quantization.

\begin{acknowledgements}
L. B. Castro acknowledges the support provided in part by funds from CNPq, Brazil, Grant No. 308172/2023-0, FAPEMA, and CAPES - Finance code 001. The authors would like to thank the anonymous reviewer for their helpful comments that have contributed to improving the quality of the paper.
\end{acknowledgements}

\bibliographystyle{spphys}       % APS-like style for physics
%\bibliography{mybibfile2020.bib}

\end{document}